\documentclass[a4paper,amsmath,amssymb,superscriptaddress,aps,prl,reprint,showpacs]{revtex4-1}
\usepackage{xcolor,graphicx}
\usepackage{subfigure}
\usepackage{MnSymbol}

\renewcommand{\vec}[1]{\mathbf{#1}}
\begin{document}
\title{Nontrivial Triplon Topology and Triplon Liquid in Kitaev-Heisenberg--type Excitonic Magnets}

\author{Pavel S. Anisimov}
\author{Friedemann Aust}
 \affiliation{Institute for Functional Matter and Quantum Technologies, University of Stuttgart,
Pfaffenwaldring 57, 
70550 Stuttgart, Germany}
\affiliation{Center for Integrated Quantum Science and Technology, University of Stuttgart,
Pfaffenwaldring 57, 
70550 Stuttgart, Germany}
\author{Giniyat Khaliullin}
 \affiliation{Max Planck Institute for Solid State Research, Heisenbergstra\ss e 1, 70569 Stuttgart}
\author{Maria Daghofer}
 \affiliation{Institute for Functional Matter and Quantum Technologies, University of Stuttgart,
Pfaffenwaldring 57, 
70550 Stuttgart, Germany}
\affiliation{Center for Integrated Quantum Science and Technology, University of Stuttgart,
Pfaffenwaldring 57, 
70550 Stuttgart, Germany}

\begin{abstract}
The combination of strong spin-orbit coupling and correlations,
e.g. in ruthenates and iridates, has been proposed as a means to
realize quantum materials with nontrivial topological properties. We
discuss here Mott insulators where
onsite spin-orbit coupling favors a local $J_{\textrm{tot}}=0$ singlet ground state. We
investigate excitations into a low-lying triplet, triplons, and find
them to acquire nontrivial band topology in a magnetic field. We also
comment on magnetic states resulting from triplon 
condensation, where we find, in addition to the same ordered phases
known from the $J_{\textrm{tot}}=\tfrac{1}{2}$ Kitaev-Heisenberg model, a triplon
liquid taking the parameter space of Kitaev's spin liquid.
\end{abstract}
\date{\today} 


\maketitle

Prime candidate systems for the interaction of spin-orbit coupling
with substantial electronic correlations are those containing $4d$
and $5d$ transition metals, where 'topological Mott insulators'~\cite{Pesin:2010ju}
or topological spin liquids were proposed. 
A prominent example is the prediction of Kitaev's spin
liquid~\cite{Kitaev:2006ik} in materials with a single hole in the $t_{2g}$
levels~\cite{Jackeli:2009p2016,Chaloupka:2013ju}. Strong research
activity has subsequently focused on honeycomb
iridates~\cite{0953-8984-29-49-493002} and on
$\alpha$-RuCl$_3$~\cite{PhysRevB.90.041112,PhysRevB.91.094422}. Encouragingly,
H$_3$LiIr$_2$O$_6$ does indeed not show magnetic
order~\cite{kitaev_H3LiRu2O6} and zig-zag order
in $\alpha$-RuCl$_3$ can be suppressed by a 
magnetic field~\cite{field_kitaev_theory,PhysRevLett.119.037201}. In
the latter case, a thermal Hall effect due to the Majorana edge 
states has been reported~\cite{thermal_hall_RuCl3_18}.

Current interest has similarly been drawn to spin-orbit coupled Mott insulators with \emph{two}
holes in the $t_{2g}$ shell. In addition to total spin $S=1$,
they would have an effective orbital angular momentum $L=1$, and spin-orbit
coupling prefers their opposite orientation into a singlet ground
state $J_{\textrm{tot}}=0$. On the other hand, magnetic superexchange between two
ions involves the excited states with $J_{\textrm{tot}}>0$. This superexchange can drive
excitonic magnetism via the condensation of bosonic 'triplons'~\cite{Khaliullin:2013du,PhysRevB.91.054412}.

While the classical limit of this scenario is governed by the same
symmetries -- and thus by similar magnetic ordering patterns -- as the
$J_{\textrm{tot}}=\tfrac{1}{2}$ scenario, the underlying degree of freedom is a
superposition of the $J_{\textrm{tot}}=0$ and $J_{\textrm{tot}}=1$ states. In addition
to opening the route to unconventional collective state like triplet
superconductivity~\cite{Chaloupka:2016gx}, this has a decisive impact on
excitations, e.g. on their dispersion in the Brillouin zone. 
With the observation of an amplitude 'Higgs' mode,
Ca$_2$RuO$_4$ has been argued to realize such a scenario close to a quantum
critical point~\cite{Higgs_Ru,PhysRevLett.119.067201}. 

Here, we investigate this scenario on the honeycomb lattice, a model
that should be appropriate to compounds like Li$_2$RuO$_3$~\cite{Li2RuO3_2007} and
Ag$_3$LiRu$_2$O$_6$~\cite{C0JM00678E}, and whose low
coordination number has been proposed to make it susceptible to states
with enhanced quantum fluctuations~\cite{Khaliullin:2013du}. We focus
first on the regime with dominant $J_{\textrm{tot}}=0$ character, i.e., where
onsite spin-orbit coupling dominates over intersite superexchange, as
found for $d^4$ iridates with a  double-perovskite
lattice~\cite{PhysRevB.93.035129,PhysRevLett.120.237204,PhysRevLett.118.086401}. We
find that excitations  become topologically nontrivial in 
magnetic fields. This implies features like protected
edge states crossing triplon-band gaps, similar to the topological magnon edge states discussed as spin
conductors with reduced dissipation~\cite{PhysRevB.87.174427,PhysRevB.91.174409}, and the thermal Hall
effect~\cite{triplon_sr_15,PhysRevB.95.195137,PhysRevB.89.134409,review_thermal_hall_17,Exp_magnon_hall_10}. 

We also present a phase diagram of the magnetic states emerging once the $J_{\textrm{tot}}=1$
states become more dominant. We find magnetically
ordered phases analogous to those of the  $J_{\textrm{tot}}=\tfrac{1}{2}$ Kitaev-Heisenberg
model, and also a disordered phase taking the place of Kitaev's spin
liquid. This `triplon liquid' realizes a
quantum-mechanical order-by-disorder scenario, where quantum
fluctuations select a unique gapped ground state from classically
degenerate dimer coverings.

\begin{figure}
  \includegraphics[width=\columnwidth]{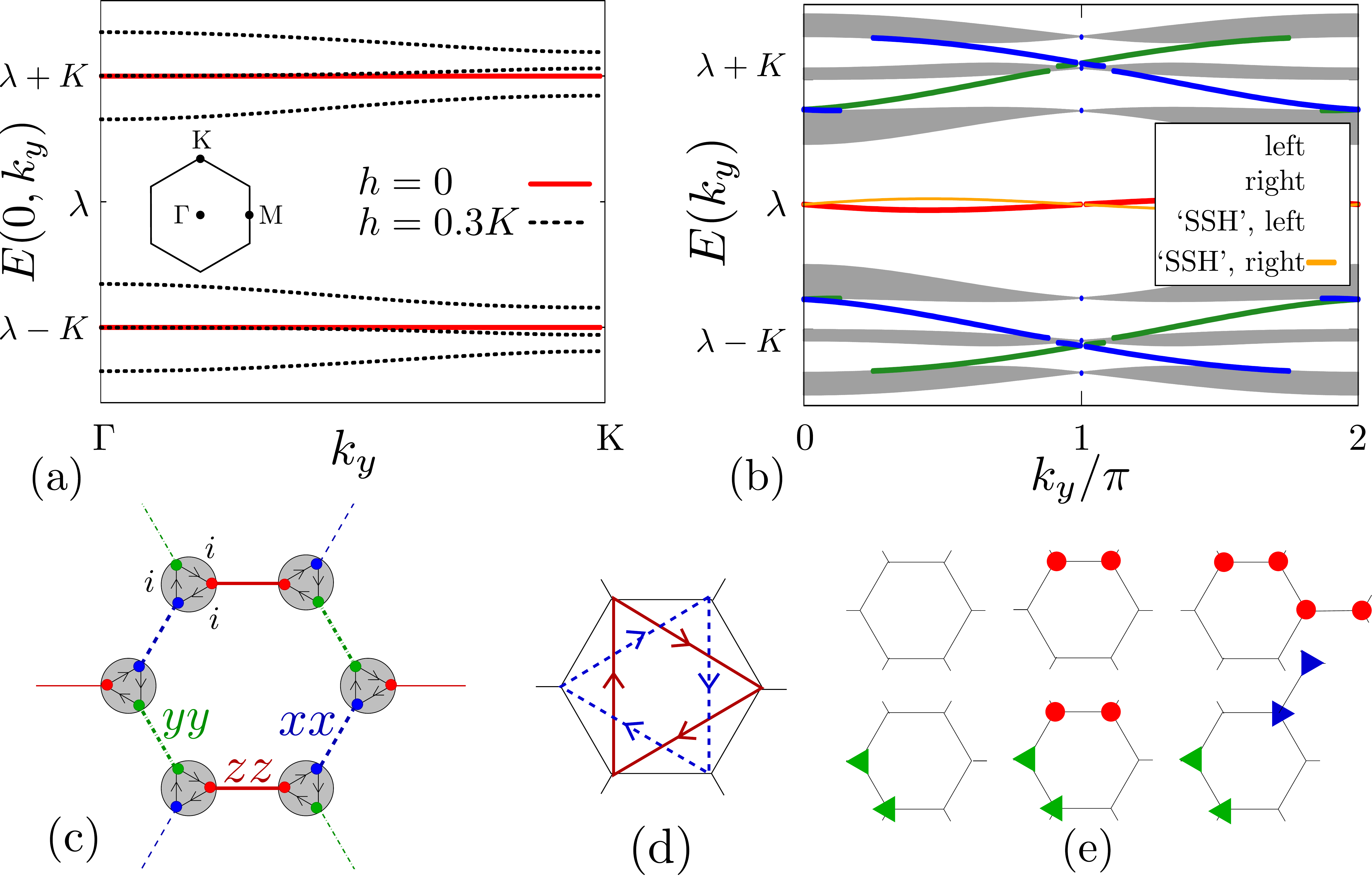}
  \caption{
    (a) Triplon bands for momentum $\vec{k}$ along the line $\Gamma=(0,0)$  to
    $K=(0,\tfrac{2\pi}{\sqrt{3}})$ for dominant 'Kitaev'
    coupling, i.e. for  $J=\Gamma=0$, and deep within the
    $J_{\textrm{tot}}=0$ regime, i.e. for $\lambda \gg K$.
    Bands are three-fold degenerate in the absence of a magnetic field
    and split for $\vec{h} = h (1,1,1)$ with $h=0.3K$.  Inset
    indicates the first Brillouin zone with three high-symmetry points. 
    (b) Topologically nontrivial bands, with Chern numbers
    $C=-1,0,1$ from the bottom to the top, and edge states along
    zig-zag edges, obtained for a cylinder. 
    (c) Decorated honeycomb lattice realized for $J,\Gamma\approx 0$ in a
    magnetic field $\vec{h}$ perpendicular to the plane. Thick colored lines
    are the bonds of the honeycomb lattice, triplons are
    confined to a bond for $\vec{h}=0$. Each shaded circle corresponds to
    one real-space site, $\vec{h}\neq 0$ allows onsite flavor
    transitions illustrated via triangles.
    (d) Next-nearest--neighbor (NNN) Dzyaloshinskii-Moriya (DM)
    interactions (\ref{eq:DM}), $D$ is positive (negative) for triplon
    hopping in the direction (opposite to) the arrows.  
    (e) Examples for triplon configurations found in the triplon
    liquid; right- and left-facing triangles and circles stand for
    $x$-, $y$-, and $z$-type triplons.
    \label{fig:cartoons_bands}}
\end{figure}

\emph{Model.} Based on Ref.~\cite{Khaliullin:2013du}, we model 
the strongly spin-orbit coupled $d^4$ Mott insulators as 
  \begin{align}\label{ham}
    H =& \lambda \sum_{i,\alpha} n_{i,\alpha}
    +J\sum_{\langle i,j \rangle} \left(\vec{T}^{\dagger}_{i}\vec{T}^{\phantom{\dagger}}_{j}
    - c_J \vec{T}^{\dagger}_{i}\vec{T}^{\dagger}_{j} +
    \textrm{H. c.}\right)\\ \nonumber
    &+K \sum_\alpha \sum_{\langle i,j \rangle{\parallel \alpha}}
    \left(T^{\dagger}_{i,\alpha}T^{\phantom{\dagger}}_{j,\alpha}
    - c_K T^{\dagger}_{i,\alpha}T^{\dagger}_{j,\alpha} +
    \textrm{H. c.}\right)\\ \nonumber
    &+\Gamma \sum_{\substack{\alpha\neq \beta\neq\gamma\\ \alpha\neq\gamma}}
    \sum_{\langle i,j \rangle{\parallel \alpha}}
    \left(T^{\dagger}_{i,\beta}T^{\phantom{\dagger}}_{j,\gamma} - c_\Gamma T^{\dagger}_{i,\beta}T^{\dagger}_{j,\gamma}
 + 
    \textrm{H. c.}\right)\;, \nonumber
\end{align}
where $T^{\dagger}_{i,\alpha}$ ($T^{\phantom{\dagger}}_{i,\alpha}$)
creates (annihilates) a triplon, i.e. a hard-core boson, with flavor $\alpha=x,y,z$ at site
$i$. These operators are collected into vectors
$\vec{T}^{\phantom{\dagger}}_{i} =
(T^{\phantom{\dagger}}_{i,x},T^{\phantom{\dagger}}_{i,y},T^{\phantom{\dagger}}_{i,z})$. The
honeycomb lattice is built of three bond directions, here likewise
labeled by $\alpha =x,y,z$ so that the triplon with coupling $K$
on a given bond bears the same index as the bond.
Energy $\lambda$ associated with creating a triplon is given by
spin-orbit coupling separating the $J_{\textrm{tot}}=0$ from the $J_{\textrm{tot}}=1$
states. Couplings $J$, $K$, and $\Gamma$ can be estimated
from second-order perturbation theory. The constants $c_J$, $c_K$, and
$c_\Gamma$ giving the relative strength of triplon hopping to pair
creation terms depend on the microscopic processes
involved. However, they are of order 1, and since we have verified that
the results presented here do not depend on their precise values, we
set here $c_J=c_K=c_\Gamma = 1$. The full model also features three-
and four-triplon terms, but as these only become relevant once the
groundstate contains an appreciable number of triplons, their
influence on triplon excitations of the $J_{\textrm{tot}}=0$ state and on its
ordering tendencies (before order sets it) are small. 
{They are
neglected here, but are shortly discussed in the supplemental material~\cite{supplmat}.}

For $90^\circ$ bond angles, dominant oxygen-mediated electron
hopping $t$ and neglecting Hund's rule, $K$ becomes $\approx -J$ so that
every triplon flavor can move on two kinds of bonds along a zig-zag
line through the honeycomb lattice~\cite{Khaliullin:2013du}. Hopping $t'$ due to direct
overlap between the $d$ orbitals leads to $K \gg J >0$; and if both
$t$ and $t'$ 
are present, $\Gamma \propto tt'$ becomes active. Further, Hund's rule
coupling promotes FM exchange~\cite{PhysRevB.91.054412}, processes
via $e_g$ orbitals might also
contribute~\cite{Khaliullin:2005PTP,Chaloupka:2013ju}, and a honeycomb
lattice can also arise with $180^\circ$ bond angles in
'dice-lattice' bilayer heterostructures~\cite{Okamoto:2013hr}. Since a large variety of
parameter combinations are possible, we treat $J$, $K$
and $\Gamma$ as material-dependent and vary them in the
present study.

\emph{Nontrivial triplon topology.}
For $\lambda \gg J, K, \Gamma$, the $J_{\textrm{tot}}=0$ state determines the ground state, but once a
triplon is excited, it can move to another site via the $T^{\dagger}_iT^{\phantom{\dagger}}_j$ terms of
(\ref{ham}). The $T^{\dagger}_iT^{\dagger}_j$ terms enter in order
$\tfrac{1}{\lambda}$, and we consequently neglect them in this
analysis of excitations deep within the $J_{\textrm{tot}}=0$ phase, see also  Ref.~\onlinecite{triplon_sr_15}.

 The bands  described by (\ref{ham}) have Chern number $C=0$, but can 
nevertheless show edge states. These can be most easily seen for
the extreme ``Kitaev'' limit $\lambda \gg |K|>0$ and $J=\Gamma=0$,
where one finds two groups of threefold degenerate dispersionless bands
at energies $\lambda \pm K$, see Fig.~\ref{fig:cartoons_bands}(a). Each  
corresponds to one triplon flavor and eigenstates are perfectly
localized on isolated bonds of the honeycomb lattice, see
Fig.~\ref{fig:cartoons_bands}(c). If a zig-zag
edge cuts all $z$ bonds along a vertical 
line, $z$-triplon states on the edge sites have no site to hop to, so
that their energy becomes $\lambda$ instead of $\lambda \pm K$, see
Fig.~\ref{fig:cartoons_bands}(b). Such states can be ascribed a 
topological origin~\cite{2018arXiv180906387J} that is related to the Zak
phase~\cite{2053-1583-4-1-015023} and to the 
topological end states of a Su-Shrieffer-Heeger (SSH)
chain~\cite{2018arXiv181008931N}. Very recently, a model supporting
such states has been argued to describe neutron-scattering data for
Ba$_2$CuSi$_2$O$_6$Cl$_2$~\cite{2018arXiv181008931N}.

The `SSH' edge states discussed above do not cross the gap between
triplon bands, are localized to isolated sites for $|K|\gg
|J|,|\Gamma|$,
and would thus not be good candidates for transport. Edge states
between bands with different Chern numbers, which 
do cross gaps and support a thermal Hall effect,  
need broken time-reversal symmetry. One possibility is a magnetic field
$H_{m} = \vec{h}\sum_i\vec{M}_i$, which couples to the
magnetic moment on site $i$,  
\begin{align}\label{eq:magn}
  \vec{M}_i = - \text{i}\sqrt{6}
  (\vec{T}^{\phantom{\dagger}}_i-\vec{T}^{\dagger}_i)  + \text{i} g
  \vec{T}^{\dagger}_i \times \vec{T}^{\phantom{\dagger}}_i\;,
\end{align}
with $g = \tfrac{1}{2}$~\cite{Khaliullin:2013du}. Again, the first
term linear in triplon operators is suppressed at large $\lambda$.

The second term in (\ref{eq:magn}), which drives onsite flavor transitions, can
as before be discussed most clearly for the  extreme ``Kitaev'' limit
$\lambda \gg |K|>0$ and $J=\Gamma=0$. Starting from the degenerate
dispersionless bands of 
Fig.~\ref{fig:cartoons_bands}(a),
a field $\vec{h} \parallel
  (1,1,1)$ [i.e.  perpendicular to the honeycomb plane]
allows transitions between flavors on each site.
Triplons are then no longer localized to a single bond and bands
become dispersive, see  Fig.~\ref{fig:cartoons_bands}(a) and (b). As 
illustrated in the cartoon Fig.~\ref{fig:cartoons_bands}(c), the
system in fact becomes equivalent to a decorated honeycomb lattice,
where topologically nontrivial bands can
arise~\cite{PhysRevB.81.205115}. As a result of the imaginary phase
i, see Eq.~(\ref{eq:magn}) and Fig.~\ref{fig:cartoons_bands}(c), the
top and bottom band of each triplet acquires a nontrivial 
Chern number $C=\pm 1$, and the two bands are connected by protected
edge states, see Fig.~\ref{fig:cartoons_bands}(b).  

\begin{figure}
  \includegraphics[width=\columnwidth]{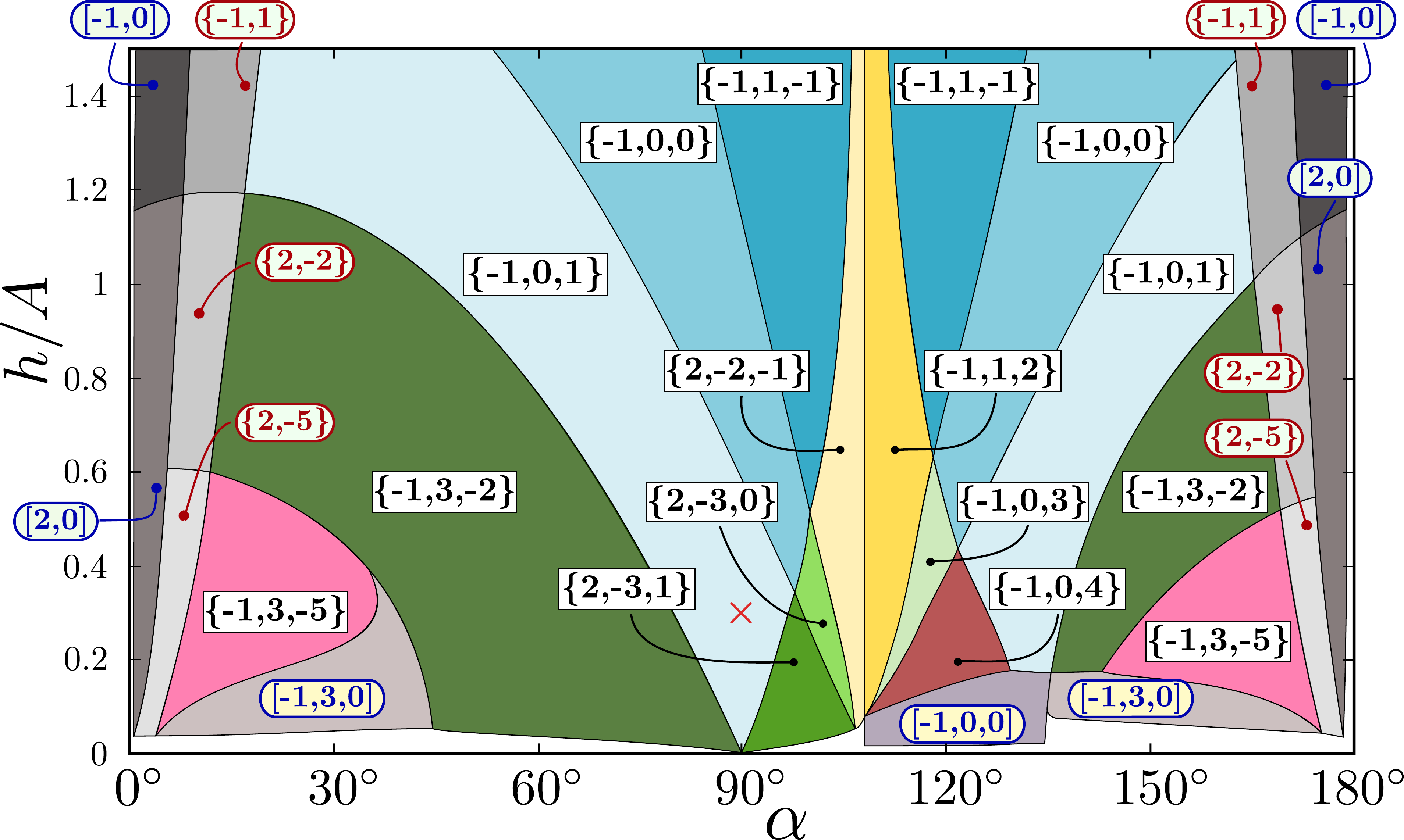}
  \caption{`Phase diagram' of the topologically nontrivial triplon
   excitations as a function of the magnetic field  $\vec{h} = h (1,1,1)$ and angle $\alpha$ defined as $J= A \cos \alpha$, $K= A \sin \alpha$. We set $\Gamma = 0$ and consider the $J_{\textrm{tot}}=0$ regime, where $\lambda \gg A$ drops out of the calculation~\cite{triplon_sr_15}. For parameter values crudely estimated to apply to transition-metal $d^4$
    systems~\cite{Khaliullin:2013du}, $h=0.1A$ corresponds to $\approx 1$ Tesla.
    The numbers given for each phase are Chern numbers: curly brackets
    refer to phases with an even number of triplon bands (four or six), where
    SSH-type edge states may additionally occur in the gap around $\lambda$. The
    Chern numbers are for the lower two or three bands, those of the
    upper two or three are opposite.  Square brackets refer to phases where
    the middle gap has closed so that there is an odd number of
    bands and no SSH-type edge states. Since our spectra are always symmetric, the middle band 
    has to have Chern number 0, the first one
    or two indices give the Chern number(s) of the band(s) below
    the middle band. The cross at $\alpha = 90^\circ$ and $h=0.3A$
    indicates parameter values used in Fig.~\ref{fig:cartoons_bands}(a,b). 
    \label{fig:topo_phases}}
\end{figure}

Figure~\ref{fig:topo_phases}  gives a phase diagram in
$J$-$K$ parameter space, with topologically nontrivial bands almost
everywhere. Gaps between Chern bands can be quite small and energy ranges of
bands may in fact overlap with indirect gaps; more robust gaps are
generally found for intermediate $\alpha$ (i.e. for large $K$).  Allowing $\Gamma \neq 0$ significantly affects
phase boundaries (not shown), but topological band character persists. 
In general, finite magnetic fields are needed, but correspond to achievable
strengths of a few Tesla for estimated parameters~\cite{Khaliullin:2013du}.
This implies that 
$t_{2g}^4$ honeycomb insulators provide a viable route to the observation
of triplon bands with Chern numbers as high as $C=5$.

Nontrivial triplon topology in coupled intersite-dimer systems arises through DM interactions~\cite{triplon_sr_15,PhysRevB.95.195137,PhysRevB.96.220405,2018arXiv180906387J},
which are symmetry-allowed on NNN-bonds and take the form:
\begin{align}\label{eq:DM}
  H_{\textrm{DM}} = \sum_{\llangle i,j \rrangle}\vec{D}_{ij} \cdot
  \vec{T}^{\dagger}_i\times\vec{T}^{\phantom{\dagger}}_j\;,
\end{align}
with $\vec{D}_{ij} = \pm D (1,1,1)$, i.e. perpendicular to the
plane;  $\llangle i,j \rrangle$ denotes NNNs and the $+$
($-$) sign applies to (anti-) clockwise motion within a 
hexagon, see Fig.~\ref{fig:cartoons_bands}(d). The similarity of 
DM term~(\ref{eq:DM}) and  
magnetization~(\ref{eq:magn}) is obvious. We have found DM
interactions to support Chern numbers $C=\pm 1$ in the absence of a
magnetic field, e.g.  for $J=1$, $\Gamma=K=0$, and
$\vec{h}=0$. However, the gaps are here rather fragile and nontrivial
band topology is lost for finite $K$ and $\Gamma$ of the order of
$D$. As NNN DM terms are in general expected to be rather
smaller than NN interactions $K$ and $\Gamma$,
this suggests a minor role for the former~\footnote{DM interactions can gain
in  importance if a trigonal crystal field 
  $\Delta\sum_{\alpha\neq\beta}T^{\dagger}_{\alpha}T_{\beta}$  
splits the $a_{1g}$ triplon mode off the $e_g'$
modes. Anisotropies due to $K$ and $\Gamma$ then average out, and the
DM interaction together with $J$ yields a perfect analogy to
Haldane's anomalous quantum-Hall-effect
model~\cite{PhysRevLett.61.2015} with Chern numbers $\pm 1$}.

\begin{figure}
  \includegraphics[width=\columnwidth]{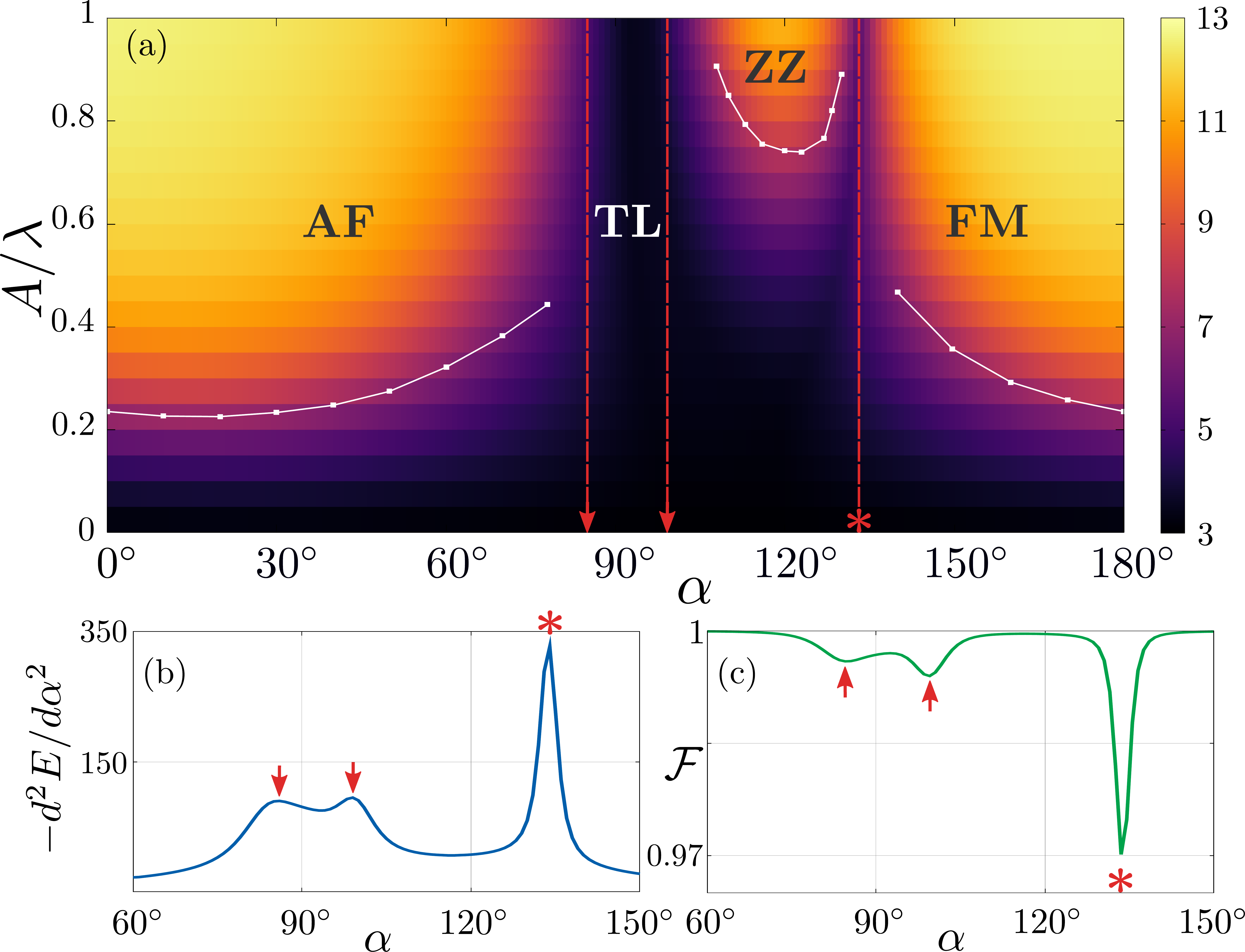}
  \caption{Magnetic phase diagram. (a) Phase diagram based on the
    spin-structure factor $S^{\alpha,\beta}(\vec{k})$, see
    (\ref{eq:sk}), obtained by ED of a 12-site 
    cluster.  Couplings are parameterized as $J=  A \cos \alpha$,
    $K=A\sin \alpha$, $\Gamma = 0$. Color refers to the maximal
    $S^{\alpha,\beta}(\vec{k})$; (AF), FM, and ZZ
    indicate (anti)ferromagnetic and zigzag order. White dots give the inflection point of
    $S^{\alpha,\beta}(\vec{k})$ vs. $A$. TL labels the region with
    appreciable triplon density ($\to 0.333$ for large $A$) but small
    $S^{\alpha,\beta}(\vec{k})$, i.e., the `triplon liquid'.
    (b) Fidelity $\langle \phi_0(\alpha) |
    \phi_0(\alpha+\textrm{d}\alpha)\rangle$  
    and (c) second derivative of the ground-state energy for
    $A=\lambda$.  Arrows and asterisk indicate the positions of the
    peaks in (b) and (c). \label{fig:ed_phases}}
\end{figure}

\emph{Magnetic Phase Diagram and triplon liquid.}
While a detailed investigation of the model's magnetic phases is
beyond the scope of this work~\footnote{An extensive ED study will be presented elsewhere, J. Chaloupka and G. Khaliullin, to be published}, we shortly discuss
their basic features. The ordering vector expected for a
magnetically ordered phase is the one where the triplon
excitations first reach zero energy. The $T^{\dagger}_iT^{\dagger}_j$
terms in the Hamiltonian have to be included here. We have
accordingly used a Bogoliubov-de\;Gennes transformation~\cite{PhysRevB.49.8901},
which neglects the hard-core constraint of the triplons, 
and exact diagonalization (ED), which is restricted to small
clusters. We additionally interpolated between these two approaches using
cluster-perturbation theory, which incorporates the hard-core constraint
within the directly solved cluster.

For $\Gamma = 0$, the phase diagram obtained from ED is given in Fig.~\ref{fig:ed_phases}(a). The dark region
in the middle corresponds to the $J_{\textrm{tot}}= 0$ regime, where hardly any
triplons are mixed into the ground state $|\phi_0\rangle$ and where magnetic structure
factors 
\begin{align}\label{eq:sk}
 S^{\alpha,\beta}(\vec{k})  = \parallel \sum_{i} \textrm{e}^{\text{i}\vec{k}\vec{r}_i}(T_{i,\alpha}^\dagger
 - T_{i,\beta}^{\phantom{\dagger}})|\phi_0\rangle \parallel^2
\end{align}
are thus small for any $\vec{k}$. The ground-state fidelity 
in Fig.~\ref{fig:ed_phases}(b) as well as the second derivative of the ground-state energy
in Fig.~\ref{fig:ed_phases}(c) have here a single peak, which
indicates a first-order phase transition.
The canonical-boson treatment and
cluster-perturbation theory agree with these phase
boundaries, which furthermore correspond more closely  to classical predictions than in the $J_{\textrm{tot}}=\tfrac{1}{2}$
Kitaev-Heisenberg model~\cite{PhysRevB.95.024426}. As in a
classical model, our phase diagram going from 0 to 180$^\circ$ (i.e. for $K>0$) perfectly repeats itself for the 
negative-$K$ part going from 180$^\circ$ to 360$^\circ$ (except that
FM and AF change places and that zigzag becomes stripy).

Differences between the classical analysis and
ED arise near the the 'Kitaev' limits
$J=\Gamma=0$. The fidelity and second energy derivative obtained from ED, see
Figs.~\ref{fig:ed_phases}(b,c), argue here against the single first-order
transition of the classical scenario and in favor of an intermediate phase in a narrow but
finite parameter regime around the Kitaev points.  
          {With a triplon density $n\approx 0.33$ at large $K$ (somewhat 
below the $n\approx 0.45$ of the ordered phases), the phase clearly
differs from the vacuum with $n\approx 0$, and we term it a  'triplon
liquid'.} We have 
found the phase  to be stable against small $\Gamma \neq 0$, its 
stability range is similar to that of Kitaev's spin liquid in the corresponding
$J_{\textrm{tot}}=\tfrac{1}{2}$ model~\cite{Rau:2013ul}. 
{The
  character of the present triplon liquid, however, differs from Kitaev's spin
liquid, as we find here no topological degeneracy.}

For $J=\Gamma=0$, the ground state
contains, in addition to the vacuum state, only configurations where
 $x$- ($y$-, $z$-) bosons sit on both ends of  $x$- ($y$-, $z$-) bonds,
see Fig.~\ref{fig:cartoons_bands}(e) for examples. This observation allows us to restrict
the Hilbert space to such dimer configurations and to obtain ground states of clusters with up to 30
sites; excitations going beyond this restricted Hilbert space can be
obtained for up to 18 sites. Based on the dimer observation, one can
moreover see that any structure factors (\ref{eq:sk}) are strictly
short range, and we find numerically no indications for bond order
either.   

Semi-classically, we expect for $J=\Gamma=0$  an
infinitely degenerate ground-state manifold of dimer coverings, with each dimer
in a superposition of `empty' and `occupied' and an 
energy of $E_c=-K/2$ per site for $K\gg \lambda$.  The triplon liquid has a
non-degenerate ground state with markedly lower energy.
This quantum-mechanical order-by-disorder mechanism
is largely mediated by the vacuum state, which is shared between the dimer
coverings and makes them non-orthogonal. 

          {The energy gap between the ground state and the rest of the
  spectrum allows the triplon liquid to survive small $J, \Gamma \neq 0$. We
  have also assessed the impact of three-
and four-triplon terms that were left out of the Hamiltonian (\ref{ham}), but
are present at sizeable triplon densities~\cite{Khaliullin:2013du}. We have
found them to leave the scenario of Fig.~\ref{fig:ed_phases} intact, i.e.,
the triplon liquid without long-range order remains as an intermediate phase
between the zig-zag and N\'eel AF phases~\cite{supplmat}.
}

\emph{Conclusions.}
We analyzed a singlet-triplet model for honeycomb compounds with a
strongly correlated and spin-orbit coupled $t_{2g}^4$
configuration, as e.g. appropriate for materials like
Ag$_3$LiRu$_2$O$_6$~\cite{C0JM00678E} and
Li$_2$RuO$_3$~\cite{Li2RuO3_2007}. 
{The latter might in fact be close to a quantum
critical point, because its magnetic state differs between powder~\cite{Li2RuO3_powder_16} and
single-crystal samples~\cite{PhysRevB.90.161110}. This would be consistent with a close
competition that is decided by the triplons's coupling to
the lattice.} For strong intersite
superexchange, we find magnetically ordered states (N\'eel, 
stripe and zig-zag AF and FM) as well as a triplon
liquid stabilized out of classically degenerate dimer coverings via a
quantum order-by-disorder mechanism.

At weaker superexchange, where the ground state is dominated by the
$J_{\textrm{tot}}=0$ state of the ion, excitations are found to be topologically nontrivial
as soon as orbital anisotropies become relevant. Topologically nontrivial triplon bands have been
proposed~\cite{triplon_sr_15} and found to agree with neutron scattering
data~\cite{edge_sl_17} for SrCu$_2$(BO$_3$)$_2$, whose ground state
consists of singlets on dimers; the discussion has since been extended
to other
geometries~\cite{PhysRevB.95.195137,PhysRevB.96.220405,2018arXiv180906387J}. Topological
triplon states in these dimer systems rely on DM  interactions, which we
found to compete with symmetric anisotropic exchange in the present
\emph{onsite}-singlet systems. Consequently, magnetic fields
perpendicular to the plane appear a more promising
route towards nontrivial triplon topology when anisotropic couplings
can be expected to dominate over DM interactions.

In addition to the $J_{\textrm{tot}}=0$ regime discussed above, topologically nontrivial excitations are expected to persist
into the FM phase, analogous to the nontrivial magnon topology in ferromagnetically polarized
states of the $J_{\textrm{tot}}=\tfrac{1}{2}$ Kitaev
model~\cite{PhysRevB.98.060405,PhysRevB.98.060404}. The AF patterns  require a more detailed symmetry
analysis~\cite{2018arXiv180705232L}, but may also harbor nontrivial magnon
bands. Finally, potential topological properties of the triplon liquid present
an intriguing question once a magnetic field renders the underlying single-triplon
states nontrivial.

\begin{acknowledgments}
We thank G. Jackeli and J. Chaloupka for many fruitful discussions. 
This research was supported by the Deutsche Forschungsgemeinschaft,
via the Emmy-Noether program (DA 1235/1-1) and FOR1807 (DA 1235/5-1).
\end{acknowledgments}


%

\newpage

\onecolumngrid
\begin{center}
\textbf{\large Supplemental Materials: Nontrivial Triplon Topology and Triplon Liquid in Kitaev-Heisenberg--type Excitonic Magnets}
\end{center}

\vspace{1em}
We present the three- and four-triplon terms of the Hamiltonian and
verify that the triplon liquid remains stable when they are included.
\vspace{2em}

\twocolumngrid

\setcounter{equation}{0}
\setcounter{figure}{0}
\setcounter{table}{0}
\setcounter{page}{1}
\makeatletter
\renewcommand{\theequation}{S\arabic{equation}}
\renewcommand{\thefigure}{S\arabic{figure}}
\renewcommand{\bibnumfmt}[1]{[S#1]}
\renewcommand{\citenumfont}[1]{S#1}



The effective Hamiltonian in terms of singlet and triplet states, 
obtained in second-order perturbation theory, can be written as 
\begin{equation}\label{fullham}
  H=\lambda \sum_{i,\alpha} n_{i,\alpha} + \sum_\alpha \sum_{\langle i,j \rangle{\parallel \alpha}} \left( h^{\alpha}_2+h^{\alpha}_3+h^{\alpha}_4 \right)_{ij}\;.
\end{equation}
The first term is spin-orbit coupling favoring the onsite singlet,
while the  sum over the nearest neighbors $\langle i,j \rangle$
parallel to the three directions $\alpha = x$, $y$, $z$ contains terms with
two, three, or four triplon operators $T^{\dagger}_{i,\alpha}$ or
$T^{\phantom{\dagger}}_{i,\alpha}$. 

When taking direct $d$-$d$
hopping $t'$ into account in addition to oxygen-mediated $t$ used in
Ref.~\onlinecite{smKhaliullin:2013du}, but still neglecting Hund's-rule
coupling and crystal fields, the bilinear contribution along a bond in
$z$ direction becomes 
\begin{align}
    h^{z}_2 =&\quad  \left(\frac{2}{3} \frac{t^2}{U} + \frac{1}{6} \frac{t'^2}{U} \right) \left(T^{\dagger}_{i,x} T^{\phantom{\dagger}}_{j,x} 
    + T^{\dagger}_{i,y} T^{\phantom{\dagger}}_{j,y}  + 
    \textrm{H. c.} \right)  \nonumber\\ \nonumber
    &  - \left(\frac{5}{6} \frac{t^2}{U} + \frac{1}{6} \frac{t'^2}{U} \right) \left(T^{\dagger}_{i,x} T^{\dagger }_{j,x}
    + T^{\dagger}_{i,y} T^{\dagger}_{j,y}  + 
    \textrm{H. c.} \right) \\ \nonumber
    &  + \frac{2}{3} \frac{t'^2}{U}  \left(T^{\dagger}_{i,z} T^{\phantom{\dagger}}_{j,z} 
    - T^{\dagger}_{i,z} T^{\dagger}_{j,z}  + 
    \textrm{H. c.} \right) \nonumber\\
    &+ \frac{tt'}{U} \left( \frac{1}{6} T^{\dagger}_{i,x}T^{\phantom{\dagger}}_{j,y}-\frac{1}{3}  
   T^{\dagger}_{i,x}T^{\dagger}_{j,y} + 
    \textrm{H. c.}\right)\;.\label{hamH2T}
\end{align}
Couplings on the other bonds are obtained by cyclic permutation.

While this Hamiltonian yields the form and symmetry of superexchange
terms, it neglects many processes alluded to in the main text,
especially Hund's-rule coupling and any orbitals beyond the $t_{2g}$
manifold. As has been frequently done in the
analogous $J_{\textrm{tot}}=\tfrac{1}{2}$ model, we thus allow the
couplings to vary rather freely.
In the main text, which discusses only the bilinear terms
(\ref{hamH2T}), this is done by assigning constants $J$, $K$, and
$\Gamma$ to isotropic, directional and flavor changing terms,
respectively. While their dependence on $t$, $t'$, and $U$ can be calculated from
second-order perturbation theory, they are instead used as free
parameters to take into account any neglected processes. 

This approach of choosing $J$, $K$, and $\Gamma$ connects easily to the literature on the
$J_{\textrm{tot}}=\tfrac{1}{2}$ model, but it is not clear how to
generalize it to the couplings arising the three- and four-boson
terms introduced below. We thus instead introduce parameters $X=\tfrac{t^2}{U}$,
$Y=\tfrac{t'^2}{U}$, and $Z=\tfrac{tt'}{U}$ that determine two- as
well as three- and four-triplon contributions. In order to allow for
processes not included in second-order perturbation theory, $X$, $Y$,
and $Z$ are then varied independently of each other and are also
allowed to become negative. 

The three- and four-boson terms can mostly be neglected when addressing
the symmetry-breaking phase transitions into the ordered states, because
triplon densities are expected to be small around a second-order
symmetry-breaking transition. However, they could affect the triplon liquid,
where they might stabilize magnetic order once triplon densities are
substantial. We thus check their impact on the phases around the Kitaev
point. Since $\Gamma\propto Z$ is not essential for this issue, we set in the
following $Z=0$ and leave the corresponding terms out for simplicity.

The three-triplon terms, except the terms $\propto Z\propto tt'$ that are left
out here, are
    \begin{align}\label{hamH3T}
      h^{z}_3 =&  \frac{1}{\sqrt{24}} \frac{t^2+t'^2}{U}  \left(T^{\phantom{\dagger}}_{i,x} T^{\dagger}_{j,y} T^{\phantom{\dagger}}_{j,z}
      - T^{\phantom{\dagger}}_{i,y} T^{\dagger}_{j,x} T^{\phantom{\dagger}}_{j,z} \right)  \\ \nonumber
      &  + \left(\frac{1}{\sqrt{6}} \frac{t^2}{U} + \frac{1}{\sqrt{24}}
      \frac{t'^2}{U} \right) \left(T^{\phantom{\dagger}}_{i,y}
      T^{\dagger}_{j,z} T^{\phantom{\dagger}}_{j,x} 
      - T^{\phantom{\dagger}}_{i,x} T^{\dagger}_{j,z} T^{\phantom{\dagger}}_{j,y} \right) \\ \nonumber
      &  + \sqrt{\frac{3}{8}} \frac{t^2}{U}  \left(T^{\phantom{\dagger}}_{i,z} T^{\dagger}_{j,x} T^{\phantom{\dagger}}_{j,y}
    - T^{\phantom{\dagger}}_{i,z} T^{\dagger}_{j,y} T^{\phantom{\dagger}}_{j,x} \right)  +  \textrm{H. c.} + i \leftrightarrow j  \;, \nonumber
    \end{align}
and cyclic permutations, the terms $\propto t^2$ can also be found in
Ref.~\onlinecite{smKhaliullin:2013du}. The four-triplon-terms can conveniently
be split into diagonal and off-diagonal parts $h_4 \equiv
h_{4,\text{diag}}+h_{4,\text{off}}$. The diagonal part  
\begin{align}\label{hamH4diag}
  h^{z}_{4,\text{diag}} =& \frac{t^2}{U} \Bigg[ 2 \left( n^{s}_{i} n^{s}_{j} +
    n^{s}_{i} n^{z}_{j} + n^{z}_{i} n^{s}_{j} +
    \vec{n}^{\phantom{\dagger}}_{i} \vec{n}^{\phantom{\dagger}}_{j} \right)
    \\ \nonumber 
    & + \frac{9}{4} \left( n^{x}_{i} n^{y}_{j} + n^{y}_{i} n^{x}_{j} +
    n^{x}_{i} n^{z}_{j} + n^{z}_{i} n^{x}_{j} + n^{y}_{i} n^{z}_{j} +
    n^{z}_{i} n^{y}_{j}  \right) \\ \nonumber 
    & + \frac{13}{6} \left( n^{s}_{i} n^{x}_{j} + n^{s}_{i} n^{y}_{j} +
    n^{x}_{i} n^{s}_{j} + n^{y}_{i} n^{s}_{j} \right) \Bigg] \\ \nonumber 
  & + \frac{t'^2}{U} \Bigg[  \frac{16}{9}  n^{s}_{i} n^{s}_{j} + 3 n^{z}_{i}
    n^{z}_{j} + \frac{7}{3} \left( n^{s}_{i} n^{z}_{j} + n^{z}_{i} n^{s}_{j}
    \right)  \\ \nonumber 
    & + \frac{3}{2} \left( n^{s}_{i} n^{x}_{j} + n^{s}_{i} n^{y}_{j} +
    n^{x}_{i} n^{s}_{j} + n^{y}_{i} n^{s}_{j} \right) \\ \nonumber 
    & + \frac{5}{4} \left( n^{x}_{i} n^{x}_{j} + n^{y}_{i} n^{y}_{j} +
    n^{x}_{i} n^{y}_{j} + n^{y}_{i} n^{x}_{j} \right)  \\ \nonumber 
    & + 2 \left( n^{x}_{i} n^{z}_{j} + n^{y}_{i} n^{z}_{j} + n^{z}_{i}
    n^{x}_{j} + n^{z}_{i} n^{y}_{j} \right) \Bigg]  \;, \nonumber 
\end{align}
is written in terms of triplon-number operators $n^{\alpha}_{j} =
T^{\dagger}_{j,\alpha} T^{\phantom{\dagger}}_{j,\alpha}$ as well as
singlet-number operator $n^{s}_{j} = 1- \sum_\alpha n^{\alpha}_{j}$.
The off-diagonal part is 
\begin{align}\nonumber
  h^{z}_{4,\text{off}} =&  -\frac{1}{4} \frac{t^2+t'^2}{U}
  \left(T^{\dagger}_{i,x} T^{\phantom{\dagger}}_{i,z} T^{\dagger}_{j,x}
  T^{\phantom{\dagger}}_{j,z} 
  + T^{\dagger}_{i,y} T^{\phantom{\dagger}}_{i,z} T^{\dagger}_{j,y}
  T^{\phantom{\dagger}}_{j,z} \right)  \\ \nonumber 
  &  - \frac{1}{4} \frac{t^2}{U}  T^{\dagger}_{i,x}
  T^{\phantom{\dagger}}_{i,y} T^{\dagger}_{j,x}
  T^{\phantom{\dagger}}_{j,y}\\ \nonumber 
  & + \frac{1}{4} \frac{t'^2}{U} \left(  T^{\dagger}_{i,x}
  T^{\phantom{\dagger}}_{i,z} T^{\dagger}_{j,z} T^{\phantom{\dagger}}_{j,x} +
  T^{\dagger}_{i,y} T^{\phantom{\dagger}}_{i,z} T^{\dagger}_{j,z}
  T^{\phantom{\dagger}}_{j,y} \right) \\  
  &  + \frac{1}{2} \frac{t^2}{U}  T^{\dagger}_{i,x}
  T^{\phantom{\dagger}}_{i,y} T^{\dagger}_{j,y} T^{\phantom{\dagger}}_{j,x}
  +   \textrm{H. c.} \;, \label{hamH4offT} 
\end{align}
and cyclic permutations.

In order to revisit the magnetic phase diagram around
$\alpha=90^\circ$  given in Fig.~3 of the
main text, we have to relate new parameters $X=\tfrac{t^2}{U}$,
and $Y=\tfrac{t'^2}{U}$ to $J$ and $K$, resp. $A$ and $\alpha$, of the main
text. We had there additionally chosen $c_K=c_J=1$, i.e., set the couplings in
front of terms like $T^{\dagger}T^{\phantom{\dagger}}$ equal to those in front
of $T^{\dagger}T^{\dagger}$. As can be seen in Eq.~(\ref{hamH2T}), this holds
exactly for contributions $\propto Y =\tfrac{t'^2}{U}$, which dominate 
near the Kitaev point, but not for those $\propto X =\tfrac{t^2}{U}$, where
the precise ratio is $\tfrac{5}{4}$. The physical meaning of using $c_J=c_K=1$
throughout is best seen by rewriting Eq.~(\ref{hamH2T}) in terms of dipolar
operators $\vec{v}=i(\vec{T}^{\dagger}-\vec{T}^{\phantom{\dagger}})$
and quadrupolar operators
$\vec{u}=\vec{T}^{\dagger}+\vec{T}^{\phantom{\dagger}}$. Setting
$c_J=c_K=1$ then amounts to neglecting quadrupolar terms, which is often
justified~\cite{Khaliullin:2013du}. Moreover, they are in fact small here, close
to the Kitaev point, and do not affect our results, as we show below. 

\begin{figure}[b]
  \includegraphics[width=0.99\columnwidth]{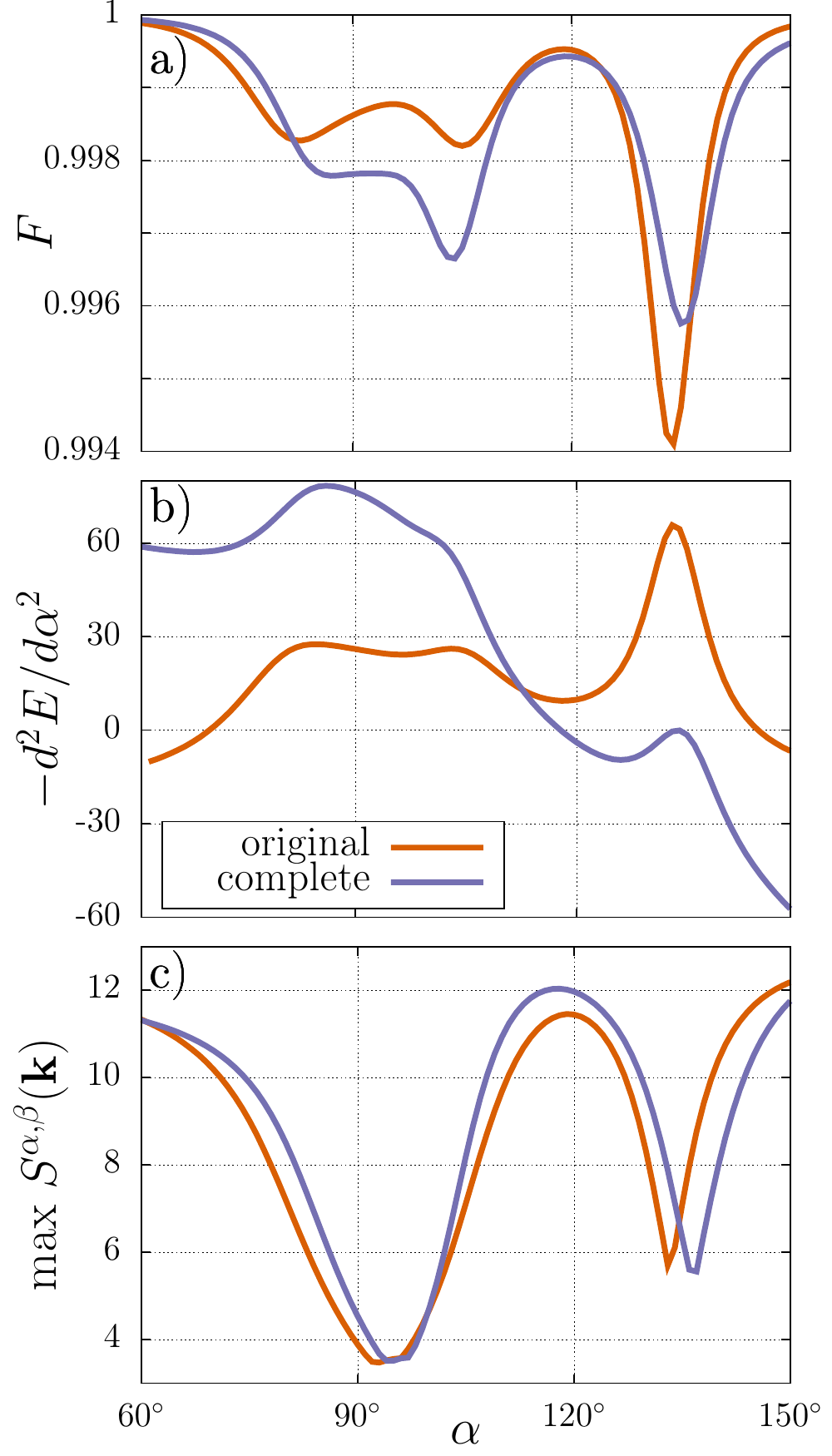}
  \caption{Impact of three- and four-boson terms on magnetic phase transitions depending on $\alpha$ for
    $A=\lambda$. Parameters $X=\tfrac{t^2}{U}$ and $Y=\tfrac{t'^2}{U}$
    of the Hamiltonian (\ref{fullham}) are determined from
    (\ref{eq:J_XY}) and (\ref{eq:K_XY}), $Z=\Gamma=0$. (a) Shows the
    fidelity (b)  the second energy derivative $\textrm{d}^2
    E(\alpha)/\textrm{d}\alpha^2$ and (c) the maximal spin-structure
    factor for full Hamiltonian (orange) and the Hamiltonian of the main text (purple),
    i.e. without $h_3$ and $h_4$
    and without quadrupolar contributions.\label{fig:phases_h3h4}}
\end{figure}

Following the above considerations, we identify $J$ and $K$ of the main
text with the isotropic and directional couplings of the dipolar fields, i.e., 
\begin{align}
  J &=  A \sin \alpha = \frac{1}{2}\left(\frac{2}{3} X + \frac{5}{6} X \right) + \frac{1}{6} Y
  = \frac{3}{4} X + \frac{1}{6} Y \label{eq:J_XY}\\
  K &=  A \sin \alpha = \frac{2}{3} Y - J = \frac{1}{2} Y - \frac{3}{4} X\;.\label{eq:K_XY}
\end{align}
Coupling parameters $X$ and $Y$ are thus obtained from $J$ and
$K$ (resp. $A$ and $\alpha$) and inserted into (\ref{hamH3T}) - (\ref{hamH4offT}) to
determine the full triplon Hamiltonian (\ref{fullham}).

Figure~\ref{fig:phases_h3h4}(a) and (b) show the fidelity end second
derivative $\frac{\textrm{d}^2 E(\alpha)}{\textrm{d}\alpha^2}$ of the
ground state energy depending on $\alpha$ and for $A=\lambda$. We see that the
substantial triplon densities allow $h_3$ and $h_4$ and the quadrupolar terms,
which are here all included, to affect the phase
boundaries to some extent. However, the interpretation of the main
text remains unchanged, i.e., we find a direct first-order transitions
between FM and zig-zag phases (at $\alpha \approx 135^\circ$) but
an intermediate regime separating AF and zig-zag phases around the
'Kitaev' point for $87^\circ \lesssim \alpha
  \lesssim 104^\circ$.

The spin-structure factor of Fig.~\ref{fig:phases_h3h4}(c) again shows
some quantitative changes, but remains qualitatively the 
same. In particular, three- and four-boson terms do not stabilize
magnetic order in this intermediate regime, the triplon liquid. The
liquid appears to be protected by the energy gap separating its ground
state from the rest of the spectrum. Additionally, a triplon density of
roughly $\tfrac{1}{3}$, even for $A\gg \lambda$ (not shown), implies
that triplons do not too often occupy nearest-neighbor sites. 

%

\end{document}